# Mg spin affects adenosinetriphosphate activity

Alexander A Tulub

Address: Department of Pure & Applied Mathematics, Saint-Petersburg State University, Universitetskaya Embankment 7/9, 199034, Saint-Petersburg, Russian Federation, Russia

Email: Alexander A Tulub - atulub@yahoo.co.uk

## Abstract

The Schlegel-Frisch ab initio molecular dynamics (ADMP) (DFT:B3LYP), $T$ = 310 K, is used to study complexation between adenosinetriphosphate (ATP), ATP subsystem, and magnesium cofactor $[Mg(H_2O)_6]^{2+}$, Mg subsystem, in a water pool, modeled with 78 water molecules, in singlet (S) and triplet (T) states. The computations prove that the way of ATP cleavage is governed by the electron spin of Mg. In the S state Mg prefers chelation of $\gamma$-$\beta$-phosphate oxygens (O1-O2), whereas in the T state it chelates $\beta$-$\alpha$-phosphate oxygens (O2-O3) or produces a single-bonded intermediate. Unlike the chelates, which initiate ionic reaction paths, the single-bonded intermediate starts off a free-radical path of ATP cleavage, yielding a highly reactive adenosinemonophosphate ion-radical, $\cdot AMP^-$, earlier observed in the CIDNP (Chemically Induced Dynamic Nuclear Polarization) experiment (A.A. Tulub, 2006). The free-radical path is highly sensitive to Mg nuclear spin, which through a hyperfine interaction favors the production of unpaired electron spins. The unique role of Mg in ATP cleavage comes through its ability to serve as a unique redox center, initially accepting an electron from ATP and then giving it back to products. Redox activity of Mg differs for T and S states and affects the number of coordinated water molecules. The findings give a new insight into the NTP (N = nucleoside) energetics and assembly of NTP-operating molecules, including proteins.

## Introduction

Without any exaggeration adenosinetriphosphate (ATP), Fig. 1, can be considered as the most important and fascinating molecule in living nature thanks to its ability to maintain the energy balance in living cells and initiate numerous biochemical reactions by releasing an imposing amount of energy, 7.3 ÷ 10.9 kcal/mol, which is then regained via recycling, a repeated process that occurs 2000 – 3000 times a day [1-6]. The energy is stored in the ATP triphosphate tail (TT), Fig. 1, and its release, as widely accepted, results from the ATP hydrolysis, partial ($\Delta E$ = -7.3 kcal/mol) or total ($\Delta E$ = -10.9 kcal/mol) [1], accompanied by the production of adenosinediphos-phate (ADP) or adenosinemonophosphate (AMP), respectively [1,5,7]. The same is true for other known in nature nucleosidetriphosphates NTPs, N = A (adenosine), C (cytidine), G (guanosine), T (thymidine), U (uridine).

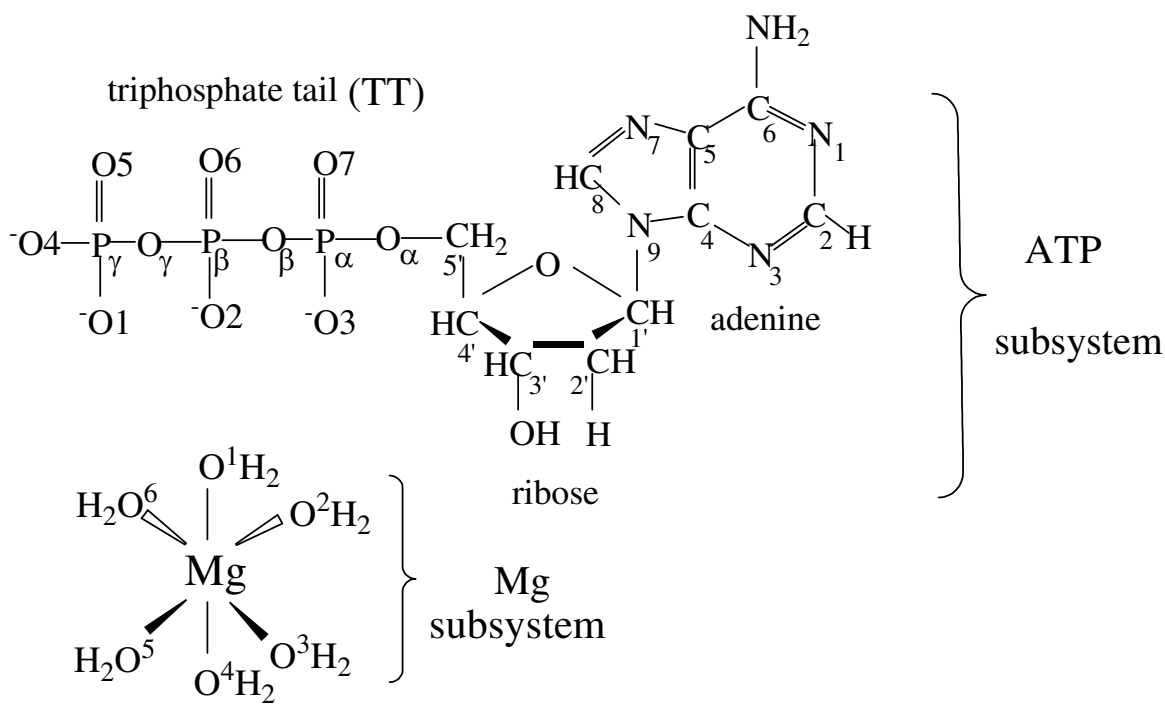

**Figure 1**
**Structure of ATP and Mg subsystems.**

In water containing environment, like living cells, ATP, and NTP in general, is quite stable, and its cleavage proceeds through overcoming a significant energy barrier of 25 kcal/mol [8-10]. The cleavage rate is highly sensitive to presence of $Mg^{2+}$ (Mg cofactor) acting on the TT as a specific catalyst [7,8]. Basically, $Mg^{2+}$ can bind to each oxygen of the TT, but generally it prefers making a chelate with O1 and O2, Fig. 1, and highly rare with O2 and O3 [1-6]. The O1-O2 chelate lowers the energy barrier by 5 kcal/mol [9,10], and the cleavage proceeds through a typical hydrolytic mechanism.

What we have just outlined is a conventional view on the NTP cleavage [1-10]. The reaction path, from its initial stage and up to final, includes only ionic forms, and diphosphate NDP and monophosphate NMP products are also ionic. The ionic forms, in turn, have paired spins (singlet states) and hardly thought to be involved in rapid reactions based on the NTP cleavage. At the same time there are numerous very rapid reactions using Mg-induced NTP cleavage. Among them are tubulin assembly into microtubules (NTP = GTP) [11,12] and DNA/RNA single chain polymerization [13-16]. These reactions can be understood only through producing short-living radical forms carrying on the unpaired spins [17]. Additionally, the ionic (hydrolytic) mechanism is unable to explain why polymerization reactions demand the presence of $Mg^{2+}$ and not other possible cations, like $Ca^{2+}$ or $Zn^{2+}$. These cations (cofactors) can easily bind the TT fragment with no serious distortions in its structure and provide similar electron density redistribution within a cluster of its atoms [1-3]. The unique role of Mg in NTP cleavage was repeatedly discussed by different groups of biologists and chemists, see [18-20] e.g., and found no appropriate answer up to present. To make the problem more puzzled, one should recall that Mg contains 10% of the active nuclear spin, $^{25}$Mg, and this spin (5/2), through a hyperfine interaction, can activate unusual reaction paths [17,21].

The paper aims to answer the following questions:

i) why Mg cation, and not other one, is essential for initiating NTP cleavage.

ii) why Mg shows different affinity to the O1-O2 and O2-O3 fragments.

iii) why chelation is a typical way of binding and why formation of a single Mg-O bond is rather atypical.

iv) why NTP cleavage is governed mostly by ionic and rarely by a free radical mechanism.

v) why Mg isotope, $^{25}$Mg, with an active nuclear spin is highly important in NTP cleavage.

To answer the above questions we have modeled the interaction between ATP (it reproduces the common features of NTP) and Mg cofactor in its singlet (S) and triplet (T) states (see the following sections). The modeling is based on the Schlegel-Frisch *ab initio* molecular dynamics (ADMP) method [22,23] with atom-centered functions [24], which logically develops the computational ideas earlier forwarded by Car and Parrinello [25].

**Model and computations**

Cellular media mostly consists of water, in which $Mg^{2+}$ exists in a water coat of octahedral configuration, Fig. 1, the most stable among other ones [26-28]. With this in mind, in our computations we consider as a realistic Mg cofactor $Mg[(H_2O)_6]^{2+}$ (Mg subsystem) interacting with ATP (ATP subsystem), Fig. 1. The geometries of both subsystems, Mg and ATP, were primarily optimized with the DFT:B3LYP (6–31G**) basis set) method. The obtained geometry parameters (interatomic distances and angles) for S states are in perfect agreement with the previously obtained ones [9,10]. T states deserve separate consideration.

Our previous computations (the computations are carried out in a cubic water box (the side length is 15.3 Å) of 132 water molecules with the full geometry optimization; DFT:B3LYP, 6–31G** basis set) show that ATP in T and S states reveal practically the same value of the total energy $E^{tot}$ [16]:$\Delta E^{tot}$(T-S) = $E^{tot}$(T) - $E^{tot}$(S) = -0.57 kcal/mol ($T$ = $0^0$ K). The difference in the $E^{tot}$ is of a *kT* level, and at $T$ = $310^0$ K (ADMP DFT:B3LYP, see below) the barrier between the T and S states is easily overcome, and both states are entangled over time. This finding implies that the T state of Mg-ATP system comes up only thanks to excitation of Mg in its water coat; the ensemble of water molecules (water pool) is always in S state.

Earlier it was shown that an isolated Mg $[(H_2O)_6]^{2+}$ (gas phase) in its T state is unstable [26-28]: the gap in the total energy between the T and S states is remarkable, $\Delta E^{tot}$(T-S) = 8.3 eV, and the complex shows the tendency to eliminate one hydrogen. Finally, it appears in the form of $[Mg(H_2O)_5OH]^+$ with the distorted geometry compared to that of the original state [29]. This is evidently not good for our purposes if we want to "prepare" the Mg $[(H_2O)_6]^{2+}$ in its T state. Fortunately, the situation is not as sad as it might seem. Our earlier computations [16] show that the Franck-Condon (F-C) S $\rightarrow$ T excitation of the Mg $[(H_2O)_6]^{2+}$ in the presence of ATP and 78 water molecules gives a much smaller gap, 0.40 eV, between the T and S states, which leaves the water molecules in the Mg shell undestroyed (the computations were carried out at $T$ = 0 K with the DFT:B3LYP method, 6–31G** basis set, and a full geometry optimization procedure at a frozen distance between the ATP and Mg atom of 8 Å; the system is fully identical to that we started off in the ADMP computations). The finding allows us to deal with triplet runs in the ADMP computations by starting from the singlet Mg $[(H_2O)_6]^{2+}$ geometry, which then, after reaching a thermodynamical equilibrium with the environment at $T$ = 310 K (see below), is converted into the triplet state through the F-C S $\rightarrow$ T excitation.

In the current molecular dynamics ADMP computations (details see below), the Mg $[(H_2O)_6]^{2+}$ and ATP (S or T state, see above) are placed in a cubic water box (the side length is 17.5 Å) of 78 water molecules, allowing each subsystem having at least three-shell water environment around them. Within the box the water molecules can change their initial positions when the Mg and ATP subsystems begin to approach. The rearrangements in the water environment during the computations, according to previous results [15,30], meet the following limits: $R_{OO} \leq$ 3.60 Å; $R_{OH} \leq$ 2.45 Å; $\phi_{OHO} \leq 45^0$ ($R_{OO}$ is the distance between oxygens, $R_{OH}$ is the distance between the oxygen and hydrogen, and $\phi_{OHO}$ is the angle between two bonds, formed by a hydrogen atom and two oxygen atoms). With these parameters the water molecules can form hydrogen bonds with the ATP on its periphery (up to 12 molecules) and between themselves in the rest water bulk, but the molecules are not allowed to approach each other close. Interatomic distances and angles within each subsystem are varied at each point of computations along the reaction coordinate. This coordinate coincides with the vector r [Mg→(O1-O2-O3)] of a length 8 ÷ 0 Å, which starts at the Mg atom (the center of the Mg complex) and goes perpendicular to the imaginary line, connecting the O1, O2, and O3 atoms, where it ends. The line has a length of ± 4.0 Å with a zero point located on the O2 atom. The named oxygens and Mg are not in the same plane. To span all the space formed by these atoms the vector r was increased in regular intervals of 0.05 Å along the imaginary line followed by its shift along the other vector n (± 1 Å) directed perpendicular to the Mg atom. Near the PES (PES = Potential Energy Surface) crossings the interval was reduced from 0.05 to 0.01 Å. In total, the ADMP DFT:B3LYP computations included 7500 independent runs.

The computations are carried out separately for S and T states with the *ab initio* molecular dynamics (ADMP) code by Schlegel and Frisch [22-24], introduced now into the *NwChem* software package (non-distributed version 5.3 plus) [31]. The method is based on the extended Lagrangian (EL) molecular dynamics approach with the wave function propagating along with the classical nuclear degrees of freedom. Unlike the widely known Car-Parrinello (CP) method [25], using EL approach and employing plane-wave basis sets (introduced into the distributed version of *NwChem*), the current method uses the atom-centered basis functions, Gaussian orbitals 6–31G** built in DFT:B3LYP [32,33,22]. Computations are carried out on a Compaq Alpha Workstation cluster, at a constant human body temperature, $T$ = 310 K. Heating to the final temperature, the initial one is 0 K, used a step-by-step procedure, each step of 10 K per 1 ps. Subsequent cooling to 0 K occurred in the reverse sequence with the same step. Such a procedure at each step provides reaching thermodynamic equilibrium between the Mg and ATP subsystems and the water pool.

Additionally, around the T-T crossing we used the same computations with a switched on hyperfine spin-coupling Hamiltonian, $S_N(^{25}Mg = 5/2) - S_e$. Charges on atoms are computed according to the Löwdin population analysis [34].

## Results and discussion

Initially separated by 8.0 Å (see section 2), the Mg and ATP subsystems in the pool of 78 water molecules immediately begin to interact by approaching each other and moving along the S or T PESs, depending on a spin state of the system. Fig. 2a displays the S and T PES projection profiles on the imaginary line, connecting the O1, O2, and O3 atoms; Fig. 2b is the 3D topside view of these PESs in the configuration space, restricted by r [Mg-(O1-O2-O3)] and O1-O2-O3 coordinates. At r [Mg-O2] = 8.0 Å the energy gap between the $T_2$ and $S_1$ states is 9.8 kcal/mol and between the $T_1$ and $S_1$ is 7.4 kcal/mol. Within all the space the T PESs go above the S ones; in addition, T and S states are remarkably separated in space. The $S_1$ PES reveals the global minimum (in our computations it is taken as a zero point, $\Delta E^{tot}$ = 0.0 kcal/mol), corresponding to the formation of a stable chelate, $[Mg(H_2O)_4-(O1-O2)ATP]^{2-}$. The distances Mg-O1 and Mg-O2 are identical, r [Mg-O1] = r [Mg-O2] = 2.05 Å, Table 1. The $S_2$ has a local minimum, which is by 21.2 kcal/

Table 1: Mg-ATP complexes of different spin and energy, $\Delta E^{tot}$ (kcal/mol), with the reference to Mg-O(1,2,3) distances, Å; in { } are the oxygens remote from the Mg.

| Complex | spin state | $\Delta E^{tot}$ | r(Mg-O1) | r(Mg-O2) | r(Mg-O3) |
|---|---|---|---|---|---|
| $[Mg(H_2O)_2$-(O2){O1}ATP] | ($T_1$ unstable) | ≥ 39.6 | | ≥ 1.90 | |
| $[Mg(H_2O)_2$-(O2){O3}ATP] | $T_2$ (unstable) | ≥ 39.6 | | ≥ 1.90 | |
| $[Mg(H_2O)_2$-(O2-O3)ATP] | $S_3$ (metastable) | 37.8 | | 1.95 | 2.04 |
| $[Mg(H_2O)_4$-(O2-O3)ATP] | $S_4$ (stable) | 34.2 | | 2.07 | 2.07 |
| $[Mg(H_2O)_3$-(O2){O1}ATP] | $S_2$ (metastable) | 21.2 | | 2.02 | 2.46 |
| $[Mg(H_2O)_4$-(O2-O1)ATP] | $S_1$ (stable) | 0.0 | 2.05 | 2.05 | |
| $[Mg(H_2O)_2$-(O2)ATP] | c (unstable) | 42.4 | | 1.87 | |

mol higher than that of the global one. The $S_2$ and $S_1$ states are separated by a low energy barrier of 1.8 kcal/mol that allows us to consider the $S_2$ minimum as a metastable, easily surmountable in real cellular environment. The $T_1$ and $T_2$ PESs, initially separated by a long-range barrier of 2.7 ÷ 2.4 kcal/mol, perpendicular to the O1-O2-O3 line and strictly positioned against the O2 atom (r [Mg-O2] = 8.0 ÷ 1.90 Å), exhibit the crossing at r [Mg-O2] = 1.90 Å, $\Delta E^{tot}$ = 39.6 kcal/mol. This crossing is responsible for producing $S_2$ and $S_3$ states and a spin-separated state, see below. The $S_3$ state reveals a local minimum at $\Delta E^{tot}$ = 37.8 kcal/mol, which then transforms into a more stable $S_4$ state, $\Delta E^{tot}$ = 34.2 kcal/mol, the $[Mg(H_2O)_2$-O2(O3)ATP$]^{2-} \Rightarrow [Mg(H_2O)_4$-O2(O3)ATP$]^{2-}$ transition. The spin-separated state, unstable and corresponding to the electron redistribution between the Mg and ATP subsystems, arises as a small reverse cone (r [Mg-O2] = 1.87 Å) of 2.8 kcal/mol height above the $T_1$-$T_2$ crossing, Table 1.

The scenario, whereby T and S PESs manifest themselves in the $\Delta E^{tot}$-r [Mg-(O1-O2-O3)]-(O1-O2-O3) configuration space, is as follows. Initially, the ATP and Mg subsystems carry the charge of -4 and +2, respectively. Upon interaction, we observe an immediate charge transfer from the ATP to Mg subsystem, r [Mg-(O1-O2-O3)] = 8.0 Å, that results in decreasing the positive charge on the Mg from +1.44 to +0.36 (S) and +0.21 (T) and making the original six-coordinated Mg water complexes unstable. Slightly charged, the Mg complexes in S and T states begin to lose a part of their initially coordinated water molecules. The loss of water molecules is not stochastic: the two remotest from the ATP water molecules ($H_2O^5$ and $H_2O^4$, Mg subsystem, Fig. 1) begin to leave the original complex one by one. This is accompanied by changing the interatomic distances and angles within each complex, but these changes are not dramatic. When the distance r [Mg-O$^5$(O$^4$)H$_2$] reaches the values of 2.60(S) ÷ 2.75(T) Å, one can observe a regrouping between the rest water molecules in such a way that opens direct Coulomb interactions between the Mg and the O1, O2, and O3 atoms ($H_2O^6$, $H_2O^1$, $H_2O^2$, and $H_2O^3$ appear to be turned away from the O1-O2-O3 line), Fig. 1. Up to this moment, the picture is practically identical to S and T states, r [Mg-(O1-O2-O3)] = 7.55(S) ÷ 7.50(T) Å, so that the changes mostly affect the Mg subsystem, leaving the ATP one practically undisturbed. This is in agreement with the previous results showing that the isolated ATP in S and T states has the same energy.

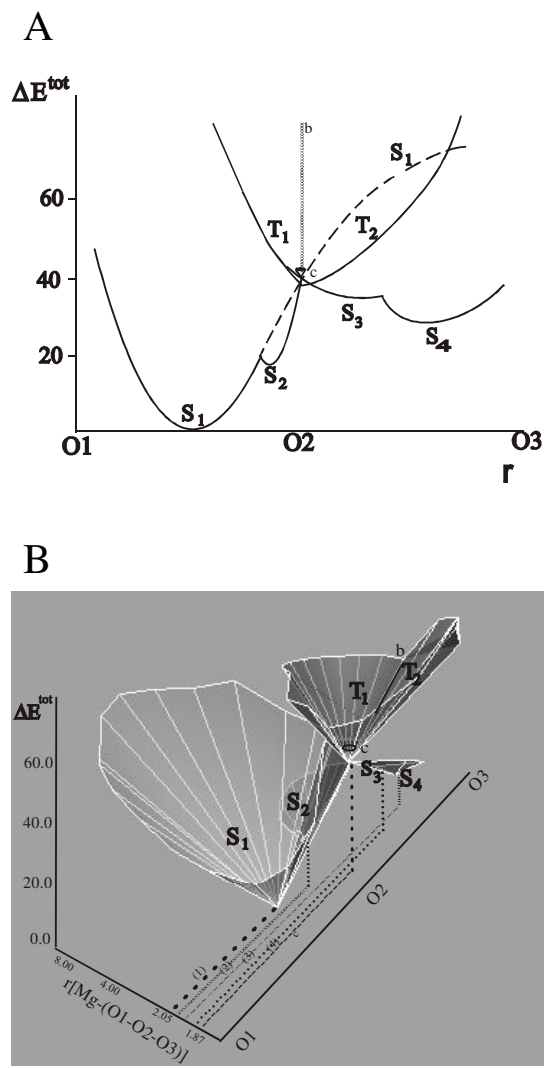

**Figure 2**
**A) S and T PES projection profiles on the imaginary line, connecting the O1, O2, and O3 oxygens of ATP.** Solid lines indicate the front PES profiles and dashed line the remote one; b, corresponds to the barrier, separating the $T_1$ and $T_2$ PESs; the cone, c, corresponds to the formation of a spin-separated, ion-radical, state. The $S_2$ and $S_3$ states are considered as metastable, see text. B) 3D topside view of S and T PESs in the configuration space, restricted by r [Mg-(O1-O2-O3)] and O1-O2-O3 coordinates; designations correspond to those in Fig. 3a. The values of r [Mg-(O1-O2-O3)] are as follows: (1) – 2.05 Å; (2) – 2.03 Å; (3) – 1.95 Å; (4) – 2.07 Å; (c) – 1.87 Å.

After the regrouping ($H_2O^5$ and $H_2O^4$ are removed from the complex by 3.7 (S) and 4.2 (T) Å, respectively), the positive charge on Mg begins to restore, but the rate and extent of this process for T and S states is different, Fig. 3. The S state gains the positive charge faster than the T one. This, in turn, produces differences in a further evolution of the system. In S state the Mg, with its higher positive charge, retains four water molecules in its coordination shell and experiences the Coulomb attraction to the O1-O2 fragment, bypassing a close approach to the O2 atom, r [Mg→O2] ≥ 2.80 Å (the arrow indicates the direction of a virtual single-bond formation). This comes from the fact that the O1-O2 fragment carries a higher negative charge than the O2-O3 one: q(O1) = -0.94, q(O2) = -0.87, q(O3) = -0.82. Approaching between the $Mg(H_2O)_4$ and ATP subsystems along the $S_1$ PES r [Mg-(O1-O2) = 7.50 ÷ 2.05 Å] finally results in forming a stable chelate, [$Mg(H_2O)_4$-(O1-O2)ATP], of S symmetry, Fig. 2a,b, Fig. 3. The intermediate product with a singly charged $Mg^+$ and $ATP^{3-}$, [$Mg+(H_2O)_4$-$ATP^{3-}$]$^S$ (r [Mg-(O1-O2-O3)] = 2.80 Å), reveals no spin localization on the Mg and ATP subsystems unlike that of T state, [$Mg^+(H_2O)_2$-$ATP^{3-}$]$^T$, see below.

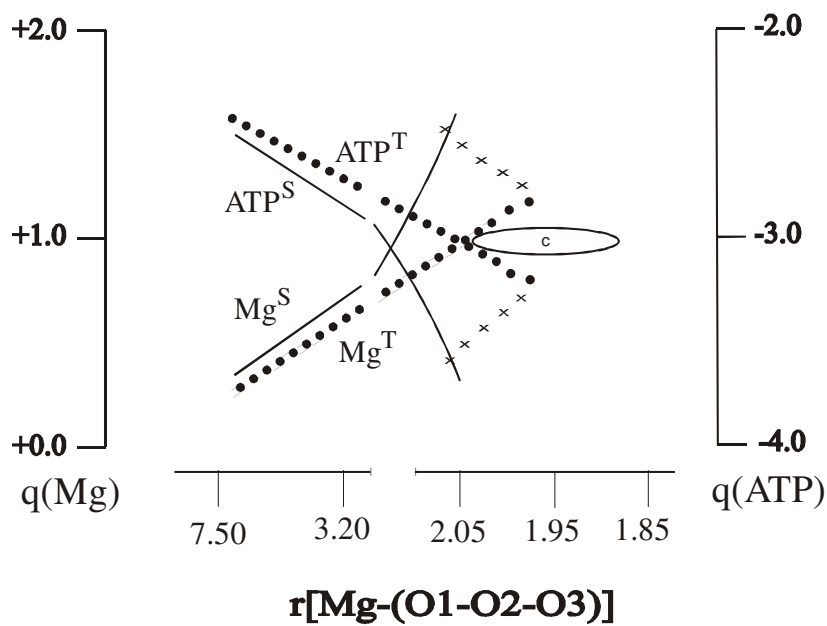

**Figure 3**
**Change in the Mg and ATP charge (q) as a function of the Mg-(O1-O2-O3) distance.** Solid lines correspond to S states and dotted ones to T states; ellipse (c) indicates the area of a spin-separated state, [•$Mg_+(H_2O)_2$-•$(O2)ATP_{-3}$]; curves marked with crosses indicate the charge reduction over the transition [$Mg(H_2O)_2$-(O2,{O3})ATP] $\Rightarrow$ [$Mg(H_2O)_4$-(O2-O3)ATP], see text.

In T state the Mg carries a lower positive charge, and the ATP terminal phosphate, $P_\gamma O_3$, appears to be slightly removed from the rest part of the ATP (r [$P_\gamma$-$O_\gamma$] is increased by 0.17 Å compared to that of isolated ATP [17] – a result of ATP perturbation from the Mg subsystem, when the two subsystems experience the Coulomb approach, r [Mg-(O1-O2-O3)] ≤ 7.50÷1.90 Å. This makes the O1-O2 fragment be lower charged compared to that of the O2-O3 one [q(O1) = -0.52, q(O2) = -0.89, q(O3) = -0.68] and makes [$Mg(H_2O)_4$]$^T$ move initially towards the O2 atom. On its way along the $T_1$ or $T_2$ PESs, the [$Mg(H_2O)_4$]$^T$ complex loses consecutively two additional water molecules in the distance interval r [Mg-(O2)] = 7.50÷2.20 Å ($H_2O^3$ and $H_2O^6$ appear to be separated from the Mg by 2.78 and 2.83 Å, respectively) and converts into [$Mg(H_2O)_2$]$^T$.

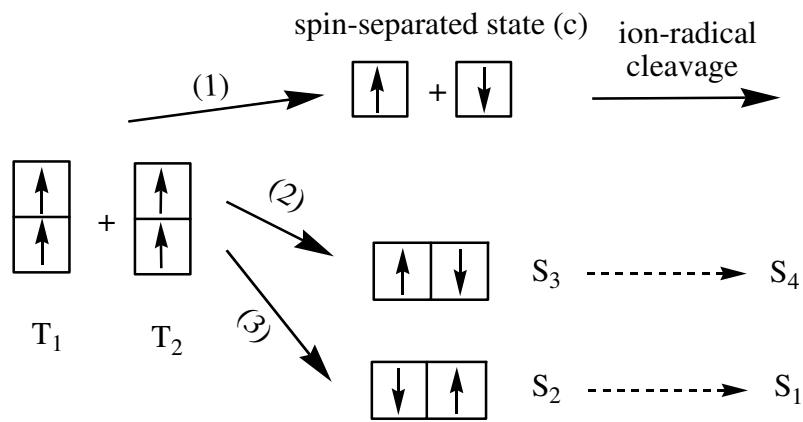

**Figure 4**
**Spin states appearing upon interaction of the $T_1$ and $T_2$ PESs around the crossing.** (1) corresponds to the production of a spin-separated state followed by ATP cleavage according to the ion-radical mechanism; (2) and (3) correspond to the production of the metastable, $S_3$ and $S_2$, states suggesting their further transformations into $S_4$ and $S_1$.

The crossing between the $T_1$ and $T_2$ PESs at r [Mg-O2] = 1.90 Å, $\Delta E^{tot}$ = 39.6 kcal/mol, deserves a separate discussion because it generates a number of states, differing in spins. The $T_1$ and $T_2$ come to the crossing having their total spins oriented in parallel, Fig. 4 (right). This is the evidence from the spin density analysis and the fact that the $T_1$ and $T_2$ PESs in the region r [Mg-(O1-O2-O3)] ≥ 1.90 Å are separated by the energy barrier, see above. The crossing state is un-stable and implies transformation into two quasi-stable S states ($S_2$ and $S_3$), Fig. 4 (left), and a spin-separated state, showing the oppositely oriented spin localization on the $Mg^+(H_2O)_2$ and $ATP^{-3}$ within the intermediate $[•Mg^+(H_2O)_2-•(O2)ATP^{-3}]$ ($\Delta E^{tot}$ = 42.4 kcal/mol), Fig. 4 (top left), Fig. 2a,b, Table 1.

The $S_2$ and $S_3$ states reveal the local minima, which could be assigned, respectively, to the production of $[Mg(H_2O)_3-(O2,\{O1\})ATP]$ ($\Delta E^{tot}$ = 21.2 kcal/mol; q [Mg] = +1.38) and $[Mg(H_2O)_2-(O2-O3)ATP]$ ($\Delta E^{tot}$ = 37.8 kcal/mol; q [Mg] = +1.22) complexes, Table 1 (in {} are the oxygens, which are remote from the Mg). The major difference in these complexes is that moving along the $S_2$ the $Mg(H_2O)_2$ adds the third water molecule and turns into a four-coordinated complex (the O1 atom is distanced from the Mg by 2.46 Å and affects the complex at a perturbation level), more stable than a three-coordinated one [35]. The $S_3$ state gains a four-coordinated stabilization by making a chelate $[Mg(H_2O)_2-(O2-O3)ATP]$ (r [Mg-O2] = 1.95 Å; r [Mg-O3] = 2.04 Å).

The $S_2$ and $S_3$ states are transformed further into more stable six-coordinated complexes: this occurs through adding (S2) the fourth water molecule (r(Mg-OH2) = 2.14 Å) and forming a stable chelate, $[Mg(H_2O)_4-(O1-O2)ATP]$ (S1 state; q [Mg] = 1.74; r [Mg-O1] = r [Mg-O2] = 2.05 Å),

Table 1, Fig. 2a,b. In the case of the $S_3$ state, the stabilization occurs through adding two water molecules with the distances $r(Mg-OH_2)$ = 2.16 and 2.23 Å. The appeared chelate $[Mg(H_2O)_4$-$(O1-O3)ATP]$ ($S_4$ state; q [Mg] = 1.66) has the identical distances r $[Mg-O_2]$ = r $[Mg-O3]$ = 2.07 Å. The $S_2 \to S_1$ and $S_3 \to S_4$ transitions proceed through overcoming small energy barriers of 1.8 and 1.4 kcal/mol, respectively.

The spin-separated intermediate $[•Mg+(H_2O)_2-•(O2)ATP^{-3}]$ is highly unstable and undergoes a rapid decomposition according to the ion-radical mechanism(*), which was closely examined by us in the previous publications [15,16]:

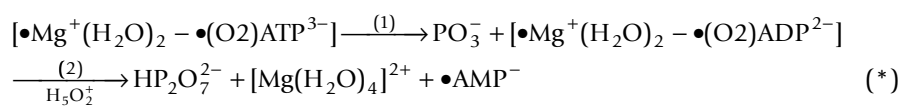

$$[•Mg^+(H_2O)_2 - •(O2)ATP^{3-}] \xrightarrow{(1)} PO_3^- + [•Mg^+(H_2O)_2 - •(O2)ADP^{2-}]$$
$$\xrightarrow[H_5O_2^+]{(2)} HP_2O_7^{2-} + [Mg(H_2O)_4]^{2+} + •AMP^- \qquad (*)$$

Reaction (*) consists of two steps: first, production of the adenosinediphosphate ion-radical, •ADP$^{2-}$, (1), bound to •Mg$^+$(H$_2$O)$_2$ within the complex $[•Mg^+(H_2O)_2-•(O2)ADP^{-3}]$, and, second, production of the adenosinemonophosphate ion-radical, •AMP$^-$, (2), upon interaction with the Zundel cation, H$_5$O$_2^+$ [36], acidic media.

The hyperfine interaction included, the barrier, initially separating the $T_1$-$T_2$ crossing and the spin-separated state (c), vanishes ($\Delta E^{tot}$ ($T_1$-$T_2$) = 39.1 kcal/mol; r [Mg-O2] = 1.87 Å). This results in the domination of (c) state over the others: $\Delta E^{tot}$(c) = 38.5 kcal/mol; $\Delta E^{tot}(S_3)$ = 40.4 kcal/mol; $\Delta E^{tot}(S_2)$ = 39.8 kcal/mol. The finding supports the idea that the active Mg nuclear spin, $^{25}$Mg, can affect the ATP cleavage according to the ion-radical mechanism [15], and the 10% content of $^{25}$Mg in natural Mg is responsible for this effect.

## Conclusion

The paper gives a new insight into the nature of NTP energetics depending on the spin activity of Mg cofactor. Up to present, Mg-induced ATP cleavage has been viewed only in the light of the hydrolytic (ionic) mechanism [1-10]. The computations show that this mechanism is not unique. Other hidden mechanisms come from the T state of Mg cofactor. When in this state, Mg cofactor directs itself to the O2-O3 fragment of NTP. This way of complexation rarely occurs, one case out of ten, but leads to significant rate enhancement in ATP cleavage, recently observed by Williams [37].

The most surprising finding is in the ability of the triplet Mg to form an unstable single-bonded state with the O2 oxygen. The state shows immediate decomposition into radical fragments – reaction (*), which starts in the femtosecond time interval [15] and finds the experimen-tal confirmation in the Chemically Induced Dynamic Nuclear Polarization (CIDNP) studies [15,16]. This Mg-O2 single-bonded state is highly sensitive to the Mg nuclear spin. If Mg is in the triplet state with a switched on nuclear spin ($^{25}$Mg), this ensures that ATP cleavage will proceed via a radical cleavage mechanism. This finding is absolutely new for molecular biology and unveils the mystery why the nature has chosen Mg, and not other cation, as an ATP/NTP catalyst. In living nature there is only 10% of Mg with an active nuclear spin ($^{25}$Mg) and 90% with inactive one. And now recall that in nine cases out of ten Mg binds to NTP through the O1-O2 atoms and only in one case through the O2-O3 atoms [1-6]! With the unchanged ratio of active to inactive spin ($R$) throughout cellular medium, a living cell, according to its program, chooses either active or inactive sort of Mg thus directing ATP cleavage through a radical or ionic path. Natural Zn has a lower content of active spin – 4.1%, and Ca is absolutely inert with its 0.15% of active spin [17]. The fact prevents these cations from entering the competition with Mg.

One might be puzzled by the fact that a relatively weak hyperfine interaction [17] can affect the way of reaction, specifically with regard to overcoming energy barriers, which normally exceed the thermal ones. If we speak about low-energy reaction paths, the effect of hyperfine coupling on overcoming chemical barriers is very unlikely. But if we are in the vicinity of conical intersections [38], where T and S states are mixed up (our case), the effect seems not surprising. The reason is that hyperfine coupling makes T and S states non-orthogonal. The nonorthogonality, in turn, produces a phase (angle) $\varphi$ between the states. In our case the T and S PESs get a rotation relative to each other by $\varphi = 15^0$. This allows the cone, Fig. 2, 3, to drop down in energy (disappear) and make the spin-separated state energetically preferable. A similar situation is observed in Watson-Crick pairing, where T and S states experience crossings [39].

The effect of hyperfine coupling on a reaction path waits for its experimental confirmation. Recently published data by Buchachenko [21] and then repeated by him in the American journals [40,41] show that replacement of nuclear inactive $^{24}$Mg and $^{26}$Mg by nuclear active $^{25}$Mg gives two-threefold enhancement in the rate of ATP synthesis. This is surely an uncommon finding, but we are not very optimistic about it. If the effect works, it should at least give tenfold rate enhancement ($R > 10$)! Moreover, with the lack of theoretical and experimental proof, confirming the existence of $^{25}$Mg$^+$ and other radical forms in the reaction, the results seem highly speculative. However, we strongly believe that our computational results will find someday reliable experimental proof

Besides nuclear specifics, distinguishing it from other cations, Mg is a unique redox center, which first accepts an electron from ATP and then gives it back to products. Ca$^{2+}$ shows the lack of this electron-based specificity and Zn$^{2+}$ is always a poor redox center. Historically, the latter was experimentally proved by Grignard [42], when he replaced Zn by Mg and thus discovered his famous reaction, which a hundred years later found a surprisingly simple explanation in S-T level conversions of Mg [43].

In summary, Mg uniquely combines redox and nuclear spin properties, and they make the cation act as a unique biological catalyst, assisting NTP to release its energy slowly (hydrolytic mechanism) or rapidly (free radical mechanism) and then gain it back via NTP synthesis. By manipulating with triplet and singlet states of Mg we can direct NTP cleavage in a desired way, and these studies are in the full swing.

**Acknowledgements**
Financial support by the *Deutsche Forschungsgemeinschaft* and *NASA Ecology Program* is gratefully acknowledged.

**References**
1.  Stryer L: *Biochemistry* 4th edition. W Freeman and Co, New York; 2002.
2.  Audesirk T, Audesirk G: *Biology, Life on Earth* 5th edition. Prentice-Hall, London; 1999.
3.  Levy C: *Elements of Biology* Addison-Wesley, New York; 1982.
4.  Nave C, Nave B: *Physics for the Health Sciences* 3rd edition. W Saunders, New York; 1985.
5.  Nelson D, Cox M: *Lehninger Principles of Biochemistry* 4th edition. W Freeman & Co, New York; 2004.
6.  Nelson P: *Biological Physics* W Freeman & Co, New York; 2004.
7.  Tuszynski J, Dixon J: *Biomedical Applications of Introductory Physics* Wiley, London-New York; 2002.
8.  Okimoto N, Yamanaka K, Ueno J, Hata M, Hoshino T, Tsuda M: **Theoretical Studies of the ATP Hydrolysis Mechanism of Myosin.** *Biophys J* 2001, **81:**.
9.  Akola J, Jones R: *J Phys Chem B* 2003, **107:**11774.
10. Grigorenko B, Rogov A, Nemukhin A: *J Phys Chem B* 2006, **110:**4407.
11. Dustin P: *Microtubules* 2nd edition. Springer, Berlin – New York; 1984.
12. Sept D, Limbach H-J, Bolterauer H, Tuszynski JA: *J Theor Biol* 1999, **197:**77.
13. Korolev G, Marchenko A: *Russian Adv Chemistry* 1999, **69:**447.
14. Odian G: *Principles of Polymerization* J Wiley & Sons, Canada; 1991.
15. Tulub AA: *Phys Chem Chem Phys* 2006, **8:**2187.
16. Stefanov V, Tulub AA: *Russian Proc Biochem & Biophys* 2007, **414:**53.
17. Nakagura S, Hayashi H: *Dynamic Spin Chemistry: Magnetic Controls and Spin Dynamics of Chemical Reactions* John Wiley, New York; 2004.
18. Ochoa S: *Enzymatic Synthesis of Ribonucleic Acid. Nobel Prize Lecture* Elsevier Publ Comp, Amsterdam; 1964.
19. Boyer P: *Energy, Life, and ATP. Nobel Prize Lecture* Elsevier Publ Comp, Amsterdam; 1997.
20. Walker J: *ATP Synthesis by Rotary Catalysis. Nobel Prize Lecture* Elsevier Publ Comp, Amsterdam; 1997.
21. Buchachenko A, Kuznetsov D: *Russian Mol Biol* 2006, **40:**12.
22. Iyengar SS, Schlegel HB, Voth GA, Millam JM, Scuseria GE, Frisch MJ: *Israeli J Chem* 2002, **42:**191.
23. Schlegel HB: *Bul Kor Chem Soc* 2003, **24:**837.
24. Iyengar SS, Schlegel HB, Voth GA: *J Phys Chem A* 2003, **107:**7269.
25. Car R, Parrinello M: *Phys Rev Lett* 1985, **55:**2471.
26. Miessler G, Tarr D: *Inorganic Chemistry* 1991.
27. Misra V, Draper D: *PNAS* 2001, **98:**12456.
28. Merrill G, Webb S, Bivin D: *J Phys Chem A* 2003, **107:**386.
29. Watanabe H, Iwata HS, Hashimoto K, Misaizu F, Fuke K: *J Amer Chem Soc* 1995, **117:**755.

30. Tulub AA: *J Chem Phys* 2004, **120:**1217.
31. *NwChem: A Computational Chemistry Package for Parallel Computers, Version 5.3 plus* non-distributed version; Pacific Northwest Laboratory, Richland, Washington; 2008.
32. Lippert G, Hutter J, Parrinello M: *Theor Chem Acta* 1999, **103:**124.
33. Schlegel HB, Millam JM, Iyengar SS, Voth GA, Daniels AD, Scuseria GE, Frisch MJ: *J Chem Phys* 2001, **114:**9758.
34. Löwdin P: *Adv Quant Chem* 1970, **5:**185.
35. Pavlenko S, Voityuk A: *Russian J Struct Chem* 1991, **32:**155.
36. Zundel G: *Adv Chem Phys* 2000, **111:**1.
37. Williams N: *J Am Chem Soc* 2000, **122:**12023.
38. Domcke W, Yarkony D, Koppel H: *Conical Intersections, Electronic Structure, Dynamics, and Spectroscopy. World Sci* 2004.
39. Tulub AA, Stefanov VE: *Chem Phys Lett* 2007, **436:**258.
40. Buchachenko AL, Kuznetsov DA, Breslavskaya NN, Orlova MA: *J Phys Chem B* 2008, **112:**2548.
41. Buchachenko AL, Kuznetsov DA: *J Am Chem Soc* 2008, **130:**12848.
42. Grignard V: *Compt Rend* 1900, **130:**1322.
43. Tulub AA, Obschei Khimii J: *Russian J General Chemistry* 2002, **76:**2219.